# Machine learning assisted global optimization of photonic devices

Zhaxylyk A. Kudyshev[1,*], Alexander V. Kildishev[1], Vladimir M. Shalaev[1], Alexandra Boltasseva[1,*]

[1]School of Electrical and Computer Engineering, Birck Nanotechnology Center and Purdue Quantum Science and Engineering Institute, Purdue University, West Lafayette, IN, 47906, USA

[*]Corresponding authors: zkudyshev@purdue.edu, aeb@purdue.edu

**Abstract:** Over the past decade, artificially engineered optical materials and nanostructured thin films have revolutionized the area of photonics by employing novel concepts of metamaterials and metasurfaces where spatially varying structures yield tailorable, "by design" effective electromagnetic properties. The current state-of-the-art approach to designing and optimizing such structures relies heavily on simplistic, intuitive shapes for their unit cells or meta-atoms. Such approach can not provide the global solution to a complex optimization problem where both meta-atoms shape, in-plane geometry, out-of-plane architecture, and constituent materials have to be properly chosen to yield the maximum performance. In this work, we present a novel machine-learning-assisted global optimization framework for photonic meta-devices design. We demonstrate that using an adversarial autoencoder coupled with a metaheuristic optimization framework significantly enhances the optimization search efficiency of the meta-devices configurations with complex topologies. We showcase the concept of physics-driven compressed design space engineering that introduces advanced regularization into the compressed space of adversarial autoencoder based on the optical responses of the devices. Beyond the significant advancement of the global optimization schemes, our approach can assist in gaining comprehensive design "intuition" by revealing the underlying physics of the optical performance of meta-devices with complex topologies and material compositions.

## 1 Introduction

Multi-constrained optimization of metamaterials [1] and metasurfaces [2–4] requires intensive computational efforts. The main goal of such optimization is to determine the distribution of constituent materials within the computational domain, which assures the best performance of the meta-device while satisfying all the constraints of the problem. Recently, various gradient-based [5–9], and metaheuristic algorithms [10,11] (evolutionary, swarm-based), have been adapted to advance nanophotonic design problems. However, even the simplest realizations of these optimization frameworks depend heavily on computationally expensive three-dimensional (3D) full-wave direct electromagnetic solvers, thus making the proposed frameworks very time-consuming and inefficient.

Moreover, the computational costs of conventional optimization methods increase with the number of additional constraints, thus making these methods less practical for highly-constrained problems. On the other hand, with the development of novel material platforms and advances in nanofabrication techniques, there is a growing interest in the multi-constrained optimization of such meta-structures, which can be decisive in addressing critical problems in the fields of space-exploration [12], quantum technology [13], energy [14], and communication [15]. There is a critical demand for efficient optimization frameworks capable of performing global optimization searches within highly dimensional parametric domains with complex optimization landscapes.

Due to its versatility and efficiency, machine learning (ML) algorithms have been successfully applied to different areas of photonics and optoelectronics. Various ML techniques have demonstrated their potential to address the bottlenecks of the conventional methods. For example, machine and deep learning models have been used in microscopy [16], quantum optics [17–19], and laser physics [20]. Recently, discriminative networks have been applied to various direct and inverse electrodynamics problems in nanophotonics [21–28]. The main advantage of the data-driven frameworks over the conventional electrodynamic simulation methods is the ability of the neural networks (NNs) to identify hidden correlations in the large datasets during the training phase and utilizing the retrieved "knowledge" to provide instantaneous solution searches, without costly computations.

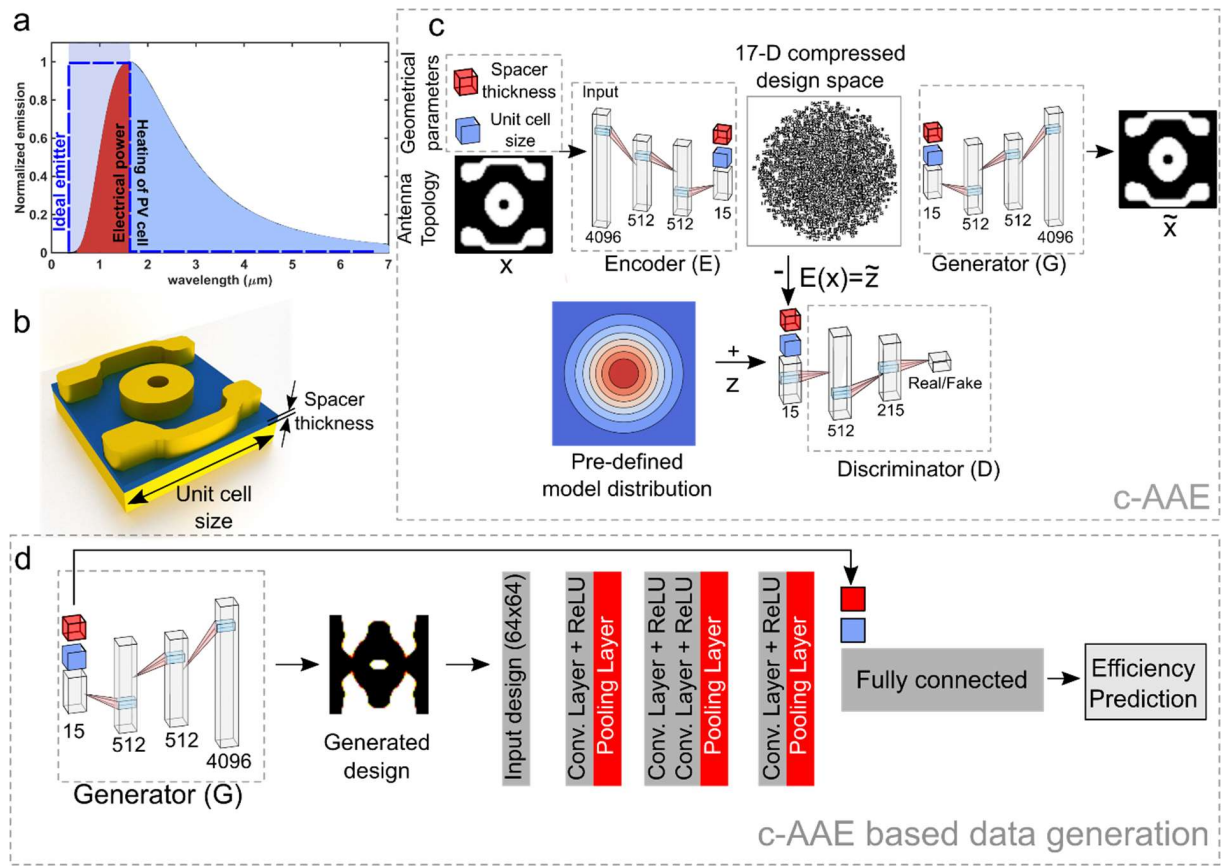

Figure 1. Conditional adversarial autoencoder based data generation. (a) Blackbody radiation of a bare heater (solid black curve) corresponding to the emission of a blackbody at 1800 °C. The grey rectangular region highlights the GaSb photovoltaic (PV) cell working band. Only in-band radiation is converted into electrical power (red area), while out-of-band radiation causes unwanted heating of the PV cell (blue area). The dark blue dashed contour corresponds to an ideal thermal emitter's emissivity/absorption spectrum. (b) The gap-plasmon thermal emitter design with a metallic TiN back reflector, $Si_3N_4$ dielectric spacer, and a top TiN plasmonic antenna. The topology optimization (TO) aims to optimize top layer with a TiN/air mixer to match the step-function emissivity pattern of an ideal emitter. (c) Training of the c-AAE network: Along with the antenna topology, the c-AAE is trained on a conditional vector with geometrical parameters of the unit cell (unit cell size, spacer thickness). (d) c-AAE-assisted rapid design generation: The trained G network is coupled with c-VGGnet for rapid efficiency estimation. The geometrical parameters of the unit cell are used as constrained labels for generation of the designs as well as estimation of the efficiencies.

Along with the optimization of geometrical parameters and prediction of the optical response of meta-structures with simplistic shapes, the advanced deep learning algorithms have been used to perform optimization of the meta-devices with complex topologies. Thus, specific classes of generative networks, such as generative adversarial networks (GANs) [29] and autoencoders [30,31], have been applied to nano-antenna design optimization. Recently, GANs and adversarial autoencoders (AAE) have been coupled with adjoint topology optimization (TO) technique for optimizing the diffractive dielectric gratings [32,33], as well as thermal emitters [34]. In [34], the authors have demonstrated that by coupling the AAE network with a conventional adjoint TO formalism, it is possible to get ~4900-time speed-up in thermal emitter optimization in comparison with conventional TO.

In this work, we have extended the AAE-assisted optimization framework and coupled it with a metaheuristic optimization engine, which allowed us to perform global optimization (GO) directly inside the compressed design space. We show that coupling the conditional AAE scheme with a differential evolution (DE) optimizer allows us to optimize the design of a thermal emitter globally, and achieve up to 96.5% thermal emission reshaping efficiency. Additionally, we demonstrate that supervised training of the c-AAE network allows us to implement physics-driven engineering of the compressed design space and helps to adapt compressed design space for better global optimization searches. Section 2 describes the main optimization problem under consideration, while section 3 introduces the main data generation framework. Specifically, section 3 describes the structure, training, and design generation process based on conditional adversarial autoencoder, as well as rapid efficiency estimation process via pre-trained conditional convolution neural network. Within section 4, we focus on the GO scheme based on the developed AAE framework. Section 5 showcases the concept of physics-driven compressed space engineering via supervised training of the conditional AAE network, as well as demonstrates the GO within the regularized compressed design space. Section 6 concludes the work.

## 2 Optimization problem

Without loss of generality, we focus on the optimization of thermal emitters for thermophotovoltaics (TPVs) utilizing GaSb photovoltaic (PV) cells with a working band ranging from $\lambda_{\min} = 0.5\,\mu\text{m}$ to $\lambda_{\max} = 1.7\,\mu\text{m}$ (shaded area in Fig. 1a). To maximize the generation of electric power from the TPV system, the emissivity of the thermal emitter should maximize in-band radiation (the red shaded area in Fig. 1a) and minimize the out-of-band radiation (the blue shaded area in Fig. 1a). Hence, the emmisivity of the ideal thermal emitter is a step function with $\varepsilon(\lambda_{\min} \leq \lambda \leq \lambda_{\max}) = 1$ and zero elsewhere (depicted as the dashed blue contour in Fig. 1a). Recently, gap plasmon structures have been proposed as a viable solution to the thermal emission reshaping problem [35–38]. However, due to the simplistic, non-optimal antenna designs, the efficiency of such thermal emitters has been limited.

Within this work, we consider a thermal emitter comprising the titanium nitride (TiN) back reflector [39], silicon nitride ($Si_3N_4$) spacer, and a TiN plasmonic antenna in the top layer (Fig. 1b). The main goal of the optimization is to determine the shape/topology of the top antenna, as well as the optimal configuration of the entire device, i.e., its 2D periodicity and spacer thickness that would drive the spectral emissivity to the step-like emissivity of the ideal emitter. Here, we do not focus on the details of the TO technique; instead, the paper is centered around the AAE-assisted global optimization framework. More details on the TO used for training set generation can be found in our prior work [34].

For assessing the performance of our designs, we define the efficiency of the thermal emitter as a product of in-band ($eff^{\text{in}}$) and out-of-band ($eff^{\text{out}}$) efficiencies as,

$$eff = eff^{\text{in}} \cdot eff^{\text{out}}, \tag{1}$$

here

$$eff^{\text{in}} = \int_{\lambda_{\min}}^{\lambda_{\max}} \varepsilon(\lambda) B_\omega(\lambda,T) d\lambda \Big/ \int_{\lambda_{\min}}^{\lambda_{\max}} B_\omega(\lambda,T) d\lambda \ ,$$

$$eff^{\text{out}} = \int_{\lambda_{\max}}^{\infty} \varepsilon_{TiN}(\lambda) B_\omega(\lambda,T) d\lambda \Big/ \int_{\lambda_{\max}}^{\infty} \varepsilon(\lambda) B_\omega(\lambda,T) d\lambda \ .$$

where the Plank law, $B_\omega(\lambda,T) = 2hc\lambda^{-3}\left(e^{hc/(\lambda k_B T)} - 1\right)^{-1}$ gives the spectral radiance of the black body at a given temperature *T* and wavelength *λ*; the fundamental constants include the Planck constant $h$, the Boltzmann constant $k_B$, and the speed of light in free space *c*. In (1) $\varepsilon(\lambda)$, $\varepsilon_{TiN}(\lambda)$ denote the spectral emissivities of the optimized emitter and a bare TiN back reflector, *T* is the working temperature of the emitter, wavelengths $\lambda_{\min}, \lambda_{\max}$ are respectively the lower and upper bounds of the PV cell operation band.

## 3 Conditional AAE for rapid design generation

To include all the design parameters into the optimization framework, we couple the conditional AAE (c-AAE) network with TO (Fig. 1c) [40]. The c-AAE network is a generative model, which consists of the encoder (E), the decoder/generator (G), and the discriminator (D). The E network is coupled with G network aiming at compression and decompression of the input design ($x$) through the so-called compressed design space. During the training phase, the E and G networks are trained to minimize the reconstruction loss between input ($x$) and the generated ($\tilde{x}$) designs by forming the 17-dimensional compressed design space. After the training process, the G network can be used to generate new designs based on the input compressed space vector $\tilde{z} = E(x)$, which is a 15D coordinate vector appended with two conditional labels $l$ (unit cell size and spacer thickness). The dimensionality of the compressed space and conditional vector is defined by the main objective of the optimization problem. It can be further enlarged according to the requirements of the problem under consideration. The regularization of the compress design space is achieved through the adversarial training process via coupling to the D network. The D network is trained to distinguish between samples from the latent distribution $q(\tilde{z})$ and the prior $p(z)$. During the training process E network is trained to reshape the compressed design space such that the D network can not distinguish between samples generated from the compressed design space and pre-defined model. The adversarial training forces the latent space to have the same data distribution as the user-defined model $p(z)$. The optimization of the c-AAE network is achieved by minimizing the following loss function:

$$L = L_{adv} + L_{rec} \tag{2}$$

The adversarial learning is achieved via a minmax game between E and D networks which aims to minimize the adversarial loss term $L_{adv}$ [40]:

$$L_{adv} = \min_E \max_D \left[\log\left(D\left(E(x),l\right)\right) + \log\left(1 - D(\tilde{z},l)\right)\right], \tag{3}$$

while the E and G are jointly trained to minimization the reconstruction loss $L_{rec}$ [40]:

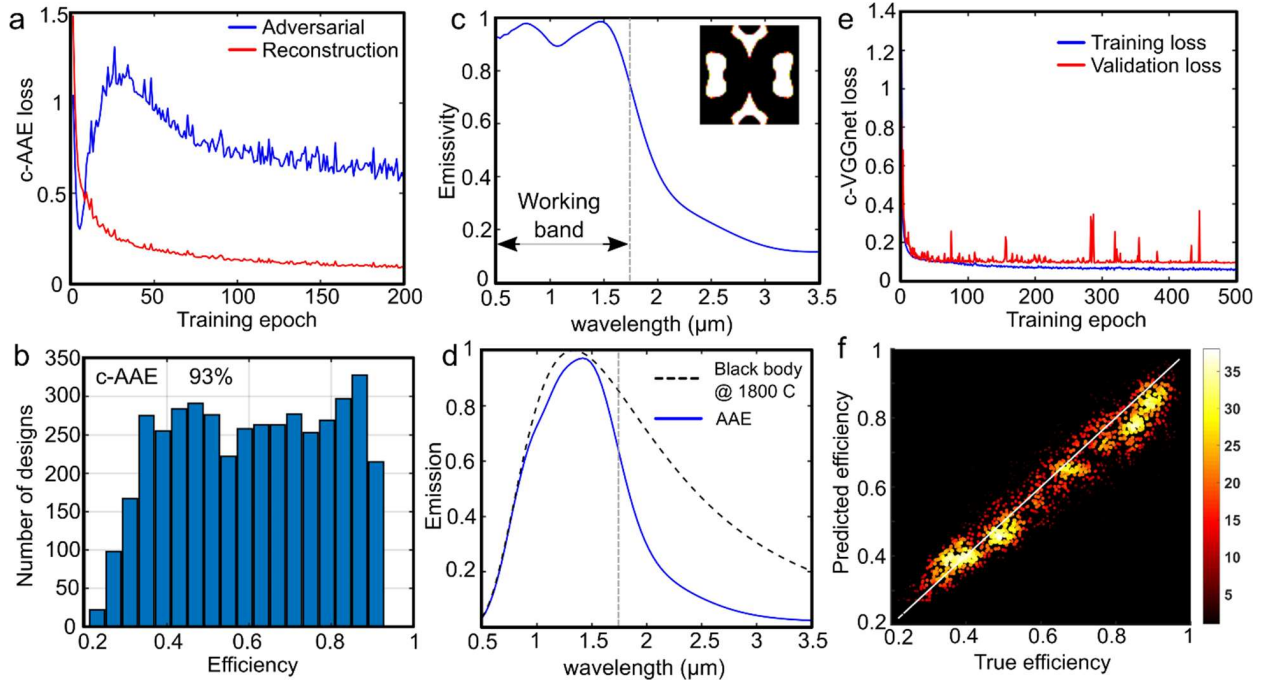

**Figure 2. Training of c-AAE and c-VGGnet networks.** (a) Evolution of the adversarial (blue line) and reconstruction (red line) losses during the c-AAE network training. (b) The efficiency distributions of the design set generated by c-AAE network via random sampling. (c) Spectral emissivity of the best design in the generated set. The dashed vertical black line shows the upped bound of the PV cell's working band. The inset shows the unit cell of the best design in the set (white color: TiN, black: air, unit cell period (x and y) 280 nm; spacer thickness, 30 nm). (d) Emission spectrum of the black body (dashed black line) and thermal emitter (solid blue line) at 1800°C. (e) Evolution of the training (blue line) and validation (red line) losses of the c-VGG predictive network during the training process. (f) the regression results performed by c-VGGnet on the testing dataset. White line shows the regression line, color map shows the number of designs tested.

$$L_{rec} = -\min_{E,G}\left[\log p\big(x\big|G(E(x),l)\big)\right]. \tag{4}$$

Once the c-AAE is trained, the G network can be used as a separate generative network that samples the new thermal emitter designs based on the input latent vector and the geometrical parameters of the unit cell (Fig. 1d). To rapidly estimate the performance of the designs, we couple the G network with a conditional VGGnet type network (c-VGGnet) [41] that estimates the efficiency of the designs based on the input binary image of the antenna, thus avoiding time-consuming full-wave simulations altogether. To realize the conditional estimation, the conditional vector is coupled to the first fully connected layer after the feature extraction part of the CNN. We give the details on the c-AAE and c-VGGnet network architecture in the supplementary materials.

The initial training set for the c-AAE network is obtained by performing TO of the thermal emitter designs for different cases of the unit cell sizes and spacer thicknesses. Specifically, the TO optimization is used to generate 100 designs for each of the period-thickness combinations: the period was chosen to be 250 nm, 280 nm, and 300 nm; and spacer thicknesses 30nm and 50 nm, thus yielding 600 designs in total. To train the c-AAE network, we use the data augmentation technique employed in [34] and increased the training set up to 24,000 designs. The periodic nature of the thermal emitters allows us to expand the initial dataset by applying lateral and rotational perturbations to the original design. Here, we expand the dataset by applying 20 lateral shifts and 90-degree rotation to each design of the original set.

The enlarged design set, as well as corresponding periodicity and spacer thickness labels, have been used to train the c-AAE network. The pre-defined model distribution is set to be Gaussian that is centered at the origin of 15D compressed space.

Figure 2a shows the adversarial (blue) and reconstruction (red) losses evolution of the c-AAE network as a function of training epochs. The reconstruction loss of the E and G coupled network decreases with the training and saturates at <0.1 value, which indicates that the G network can reconstruct input design correctly from the compressed design space. The adversarial loss decreases at the beginning of the training, which corresponds to the fact that the E network fails to generate the desired compressed space distribution at the early steps of the training. However, with the increasing number of the training epochs, adversarial loss saturates at relatively high values (~0.8), highlighting the ability of the E network to "fool" the D network and passing the sample from the compressed space as "real" through it. This fact indicates that the constructed compressed design space has data distribution close to the pre-defined one. Once the training of the c-AAE network is done, we have generated 4500 thermal emitter designs via random sampling of the compressed space coordinates. Specifically, the 300 designs for each of the combinations of the periodicity (from 200 nm to 280 nm with 20 nm step) and spacer thicknesses (30 nm, 40 nm, and 50 nm) have been generated. The efficiency of each design has been assessed by FDTD simulation (Lumerical FDTD solver). Figure 2b shows the statistics of the generated dataset. The best design in the set has 93% efficiency of thermal emission reshaping and corresponds to 280 nm unit cell size and the 30 nm spacer thickness. Figures 2c and 2d, show the corresponding absorption/emissivity spectrum, as well as the gray body emission at 1800°C from the best design in the set (blue curves). The inset in Fig. 2c shows the antenna design. The complex topology of the c-AAE antenna design enables 94% mean in-band absorption and substantially suppresses the out-of-band absorption spectrum (23% mean out-of-band absorption). This absorption behavior of the thermal emitter leads to its high in-band emission and the rapid decay of the out-of-band emission.

To perform the filtering of highly efficient design as well as to avoid time-consuming full-wave analysis, the c-AAE network is coupled to the c-VGGnet regression network, which estimates the performance of the design based on the input design and unit cell parameters (Fig. 1d). The c-VGGnet regression network is trained on the design set, which is generated by the c-AAE network (Fig. 2b). This training is done due to the high designs' variance, and a broader range of the unit cell parameters vs. the TO generated set that allows for more efficient training of the c-VGG network. Employing the generated design set, unit cell parameters, and corresponding efficiency values as ground truth, the c-VGGnet is trained by using mean absolute percentage error loss function. 80% of the designs are used for training, while 20% is used for the validation loss estimation. The loss evolution during the training process is shown in Fig. 2e. The figure demonstrates that the regression loss decreases and saturates at 9%, indicating that c-VGGnet is capable of retrieving the efficiency values based on the binary image of the design with high accuracy. For assessing the performance of the regression network, we calculate the coefficient of determination $r^2$ that quantifies the ability of the network to predict the variance of the true data. While in the ideal case $r^2$ should be equal to 100%, a sufficiently high value of $r^2$ ( 87%) is achieved.

Figure 2f shows the dependence of the predicted values by the network vs. the true values. In the ideal case, the point on the colormap should coincide with the regression line (white, solid line). The integrated scheme of the c-AAE generator with c-VGGnet opens the possibility to perform rapid prototyping and efficiency estimation of the meta-devices (Fig. 1d). This combination is a crucial step for the realization of various, ML-assisted GO schemes for highly-constrained problems. The next two sections highlight one of the possible c-AAE assisted GO. The proposed approach can be integrated with the other metaheuristic and/or gradient-based optimization frameworks. The next section describes the c-AAE based GO technique based on differential evolution optimization.

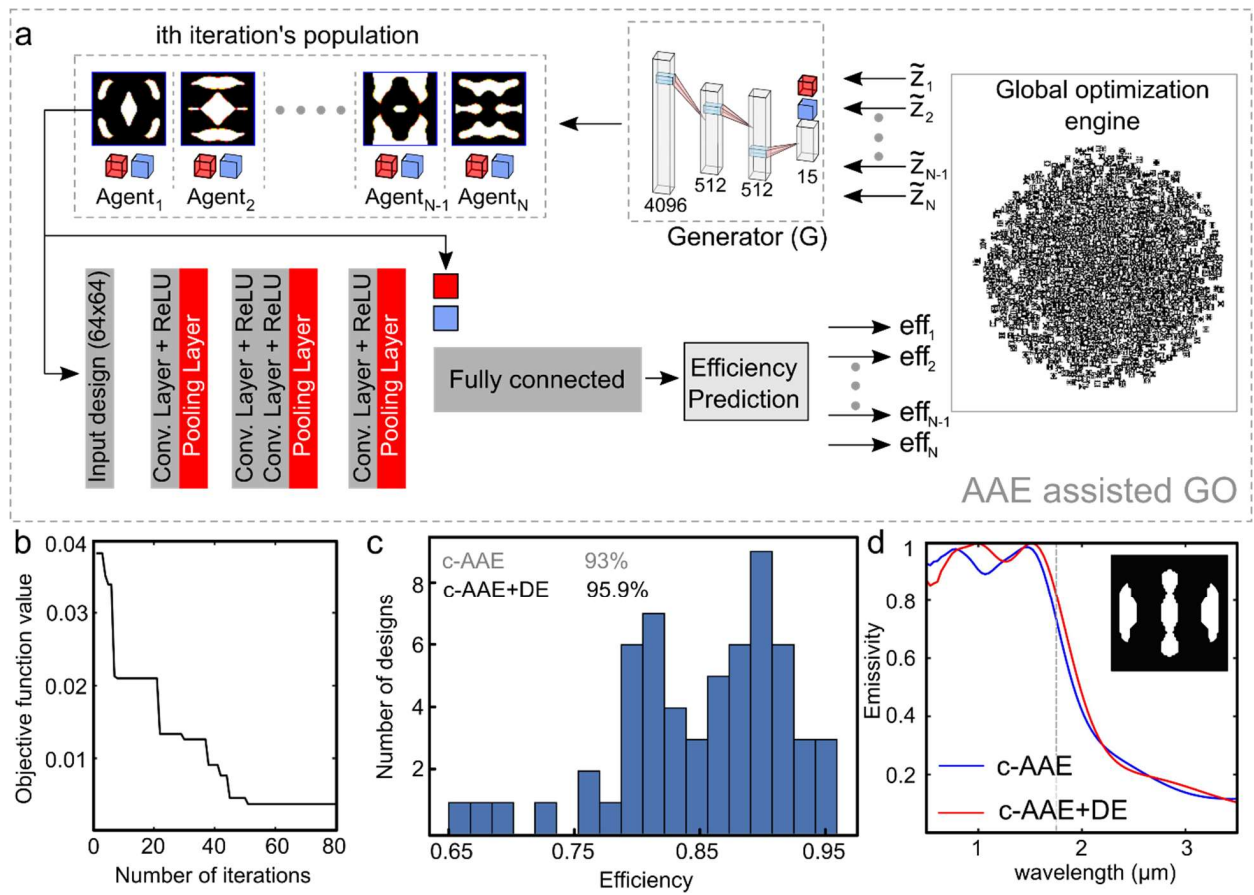

Figure 3. c-AAE-assisted global optimization. (a) Scheme of the c-AAE based GO algorithm. The GO engine is used as a black box that generates the agents' coordinates within the compressed design space and pass them to the generator of the c-AAE network. The generator samples the design, and the corresponding efficiencies are rapidly assessed via the c-VGGnet network. These efficiencies are returned to the GO engine to update the positions of the agents at the current iteration. (b) Conversancy plot of the c-AAE+DE optimization framework. (c) Efficiency distribution for 60 designs generated by the c-AAE+DE. The legend indicates the maximum efficiencies obtained via random search with in the compressed design space (93%, gray) and by GO (95.9%, black). (d) Spectral emissivity/absorption of the best designs generated by random search (blue) and GO (red). Thin vertical dashed line shows the upped bound of the PV cell's working band. The inset depicts the best unit cell design in the set (white color: TiN, black: air; unit cell period (x and y), 190 nm; spacer thickness, 45 nm).

## 4   AAE-assisted global optimization

Due to the non-convex nature of many optimization problems, the design spaces usually correspond to the rapidly changing figure of merit (FOM) landscapes that make the brute-force approach inefficient in the quest for the global solution. This issue becomes even more significant for multi-constrained problems since the exhaustive search within highly-dimensional parametric landscapes would be extremely resource heavy with no guarantee of retrieving the most optimal solution. Hence, it is crucial to develop a global optimization framework that is capable of using the "best of both worlds": (i) efficiency and scalability of the c-AAE based optimization framework and (ii) the ability of metaheuristic algorithms to perform global optimization searches most efficiently. To address the issues mentioned above, we develop c-AAE-assisted GO scheme, which is capable of performing global search directly within the compressed design space.

Within this work, the differential evolution(DE) algorithm [42] has been coupled to c-AAE based data generation approach for retrieving the global maximum inside the compressed design space. The DE framework is a population-based metaheuristic algorithm, which uses multiple agents to probe the solution space and evolutionally converge to global extremum. At optimization step, the positions of each of the agents in the population are updated by adding the weighted difference between two randomly selected agents from a given population to the agent with the best efficiency at the current iteration:

$$\tilde{z}_n^{i+1} = \tilde{z}_{best}^i + F\left(\tilde{z}_{r1}^i - \tilde{z}_{r2}^i\right) \quad (5)$$

Here, $\tilde{z}_{best}^i$ is a compressed design space coordinate of the agent with the best performance at $i^{th}$ iteration, $F$ is a mutation parameter, $r_{1,2} = \text{rand}(1,N)$ are the random indices, and $N$ is a total number of agents within a given population.

The coordinates of the agents at the next optimization step are updated with (5) or kept unchanged. This choice is made with a binomial distribution by generating a random number, $r = \text{rand}(0,1)$, and comparing it vs. a pre-defined recombination constant.

Within the developed optimization framework, the DE optimizer sends the set of compressed design space vectors $\tilde{z}_{1,N}^i$ to the c-AAE generator. The G network generates the designs, and the c-VGG-net estimates the efficiencies $\textit{eff}_{1,N}^i$ of each design in the set. Once these efficiencies are sent back to the DE optimizer, the coordinates of the agents are updated at the next optimization step employing the algorithm shown above. At the end of the DE optimization run, we use the best $\tilde{z}_{best}^{final}$ to generate the antenna design and retrieve the corresponding unit cell configuration. For this, we take two last elements of $\tilde{z}_{best}^{final}$ corresponding to the unit cell size and spacer thickness, encoded during the constrained training of the c-AAE network. The c-AAE based GO framework assures an extremely flexible framework that addresses highly-constrained optimization problems by enlarging the compressed design space with a larger number of conditional labels during the c-AAE training. Most importantly, the c-VGGnet regression network removes the need for time-consuming full-wave simulations at the optimization search stage. Additionally, since the proposed approach uses the global optimizer as a black-box, the developed c-AAE-assisted framework can be coupled to any global optimization techniques.

The DE optimization is implemented using the SciPy library [43]. The total population size of the DE optimizer is set to 20 agents, with a maximum iteration number of 80. The mutation and recombination coefficients are set to 0.5 and 0.7, respectively. The optimization objective function aims at minimizing the value of $1 - \textit{eff}$. Figure 3b shows the typical conversancy curve of the c-AAE assisted DE optimization (cAAE+DE). The DE optimization starts with the relatively high value of the objective function. However, with the evolution of the optimization, the DE algorithm converges to better efficiencies. The optimization stops when the objective function gets saturated, or the maximum iteration is reached. We have performed 60 c-AAE+DE optimization runs. Figure 3 c depicts the statistics of the efficiencies obtained from the runs. The analysis shows that the c-AAE+DE approach assures much better performance in comparison with the set generated directly from the c-AAE network. The mean efficiency of the distribution is 85%, while the best-obtained design has 95.9% efficiency vs. 93% of the c-AAE set (Fig. 2b). Figure 3d shows the absorption spectra of the best design in the set (red line), while the blue curve indicates the efficiency of the best design generated directly from the c-AAE network (93%). The inset illustrates the best design generated with the c-AAE+DE approach. The c-AAE+DE optimization framework leverage on the connection of the shape/topology of the metadevices and their optical responses. However it is highly desireabel to construct the physics-driven compressed design space via incorporating avalibale knowledge on the devices performance into the training process. By doing so it is possible to regularize the compressed design space for more efficient GO searches. The next section describes the physics driven compressed space engineering framework and demonstrates the performance of the GO within such design domains.

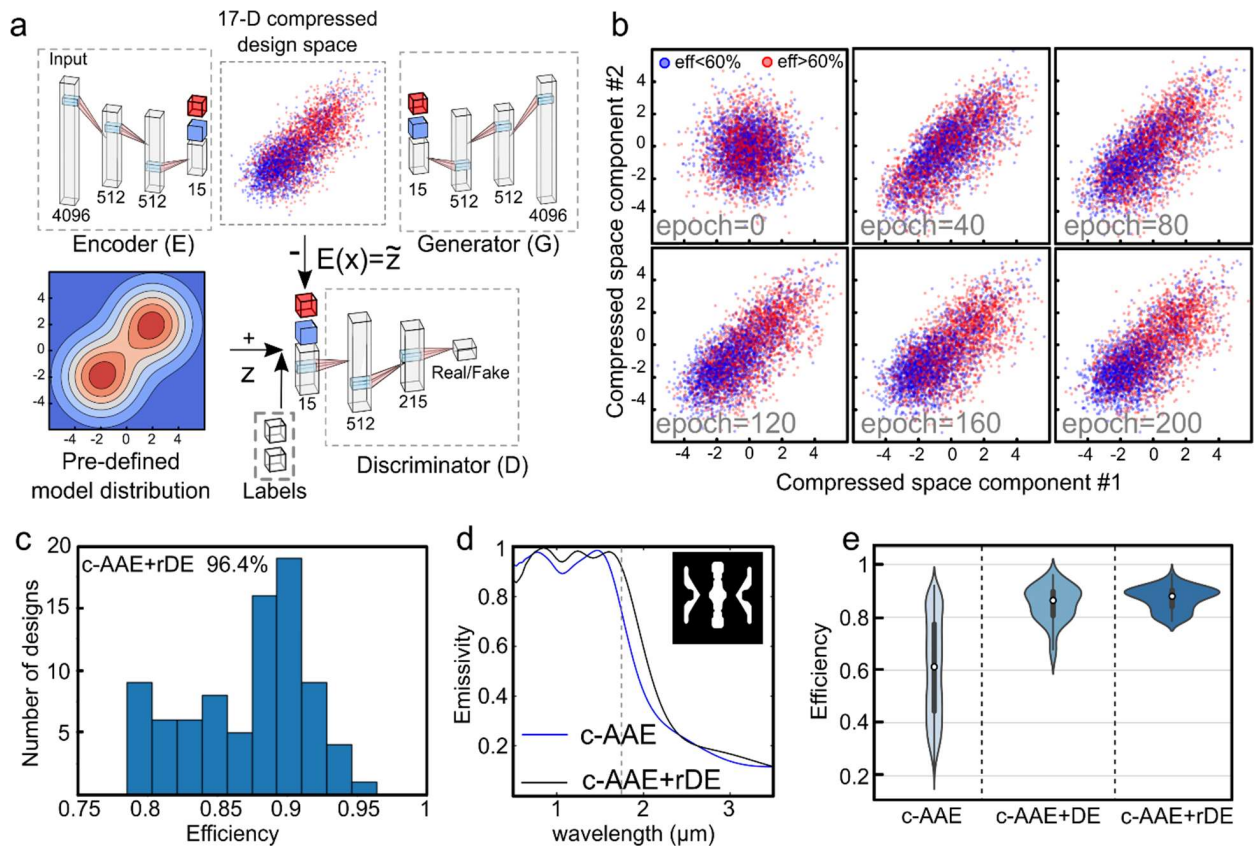

Figure 4. Physics-driven compressed design space engineering for GO. (a) Scheme of the supervised training of the c-AAE network for physics-driven compressed space construction. The predefined model is set to be a combination of two 15-D Gaussian distributions symmetrically shifted from the origin. The additional "hot" binary vectors are used during the training for regularization of the compressed space during training. (b) Evolution of the compressed design space within the training process. The scatter plots show the distribution of the designs used in the training within 2D plane cut along the first two coordinates of the 17D compressed space. The blue markers depict the LE class data, red markers show the distribution of the HE class. (c) Efficiency distribution of 80 designs globally optimized with the c-AAE+rDE. (d) Spectral emissivity/absorption of the best designs generated by random search (blue dashed line) and c-AAE+rDE (black solid line). The thin dashed vertical line shows the upped bound of the PV cell's working band. The inset shows the unit cell of the best design in the set (white color: TiN, black: air, period (x and y) 290 nm, spacer thickness 45 nm). (e) The "violin plot" of three design sets generated by, (i) random search within un-regularized compressed design space (left gray pattern, c-AAE), (ii) DE optimization within the un-regularized compressed design space (center light blue pattern, c-AAE+DE) and (iii) DE optimization within the regularized compressed design space(right dark blue pattern, c-AAE+rDE).

## 5  Physics-driven compressed space engineering

The training of the c-AAE network on TO dataset leads to the compressed design spaces, constructed based on geometrical features of the metadevices, omitting the available information regarding the optical responses of the designs. However, using available information about the essential physics of the meta-devices, it is possible to pre-engineer the compressed design space for improved performance of the GO search. Such regularization of the compressed design spaces can be introduced by choosing the physics-driven pre-defined model of the c-AAE network connected with the FOMs of the meta-devices.

This technique can be realized based on the supervised training of the c-AAE network by adjusting the pre-defined model and passing an additional binary regularization vector into the D network at the training stage (see, Fig.4a). For demonstrating the physics-driven compressed space engineering, we use two 15D Gaussian distributions in a pre-defined model and introduce an additional 2D hot vector as a label to the D network (Fig. 4a). In more detail, the designs in the training dataset are divided into two classes based on their efficiencies, (i) high efficiency (HE, eff > 60%) and (ii) low efficiency (LE, eff < 60%) classes. The prior sampling of the c-AAE model has been defined as:

$$z_i = \begin{cases} \mathrm{N}(\mu = 2, \sigma^2 = 1), & \mathit{eff}_i > 60\% \\ \mathrm{N}(\mu = -2, \sigma^2 = 1), & \mathit{eff}_i < 60\% \end{cases}, \qquad (6)$$

here $z_i$ is a sampling of the pre-defined model for $i^{th}$ design in the set, $\mathrm{N}(\mu, \sigma^2)$ is random normal distribution with mean $\mu$ and variance $\sigma^2$.

With this approach, we construct the compressed design space, with two clusterization regions separated in accord with the two efficiency-level classes. By applying the DE optimization within the HE region, we can obtain the thermal emitters with better efficiencies. Figure 4b shows the evolution of the compressed design space throughout the training process. The figure shows two first coordinates of the 17-D compressed design space. The red markers correspond to the HE class, while the blue ones represent the LE designs. Initially, all designs are sampled as a mixture of HE and LE classes. With the training, the clusterization of both classes progressively appears. The final state of the compressed design space is shown in Fig. 4b (epoch=200). Once the training is done, the DE optimization is applied to the HE designs' region of the compressed space. We perform 80 runs with c-AAE+rDE optimization technique. Figure 4c shows the efficiencies statistics of the optimized designs. The additional regularization leads to better performance of the DE optimizer vs. the un-regularized case (c-AAE+DE). Thus, the c-AAE+rDE ensures 87% mean efficiency, with the best design in the set providing 96.4% efficiency of thermal emission reshaping. Figure 4d shows the emissivity spectrum of the best design in the set. The optimized design delivers 96% of in-band emissivity and substantially suppresses the out-of-band emission. The inset shows the design of the best thermal emitter in the set.

Figure 4e shows the back-to-back comparison of the efficiencies distributions of three c-AAE based optimization frameworks, (i) the set generated from c-AAE by random search (gray), (ii) c-AAE+DE (light blue) case as well as (iii) c-AAE+rDE (dark blue). Sampling of the emitter designs from the random search (the c-AAE network with no post-selection) leads to a broad range of the efficiencies with a mean efficiency of 61% and a maximum efficiency of 93%. It can also be seen that the efficiencies are uniformly distributed along the entire range (almost uniform width of the shadowed area, left subplot in Fig. 4e). In contrast, the GO performed inside the un-regularized compressed design space enables the design generation with much better efficiencies distribution with data concentrated within [65%, 95.9%] efficiency range and the maximum designs sampled around 90% efficiency (light blue, central subplot in Fig. 4e). The regularization of the compressed space coupled with the GO search leads to even better efficiencies distribution within [79%, 96.4%] (dark blue, right subplot in Fig. 4e). This analysis clearly shows that the physical regularization of the compressed design space allows adapting the design space configuration to perform a better GO search.

# 6 Conclusion

In conclusion, we have developed a global optimization framework utilizing a c-AAE network that can be applied to a wide range of highly-constrained optimization problems in nanophotonics and plasmonics as well as in biology, chemistry, and quantum optics. We show that by applying the differential evolution optimization directly to the compressed design space, it is possible to achieve efficient optimization of the metadevices with complex topology. We numerically demonstrate advanced compressed space engineering by utilizing physics-driven regularization of the compressed design space via supervised training of the c-AAE network. The proposed physics-driven design space compression leads to significant improvement in the GO search. We also show that physics-driven regularization of the compressed design space leads to a more intuitive way of performing the GO search within the compressed space, which in turn, leads to the almost perfect performance of the optimized metadevices.

Pre-engineering of the compressed design spaces of meta-structures can be used in combination with diverse ML algorithms such as principal component analysis [44] and cluster analysis [26] both to retrieve the best possible solution of the problem and to gain hidden knowledge about the physics of the meta-structure with complex topologies. For example, analyzing the eigenmodes of the structures sampled from the high-efficiency cluster, it is possible to gain additional intuition regarding the electrodynamic mode components that lead to the optimal metadevices. This technique would allow us to generalize the physics requirements to the device design for achieving the best possible performance and reconstruct the antenna designs based on the first principles approach to the problem.

ACKNOWLEDGMENTS

The authors acknowledge partial support from the NSF ECCS award "Machine-Learning-Optimized Refractory Metasurfaces for Thermophotovoltaic Energy Conversion", Army Research Office MURI award no. W911NF-19-1-0279 and DARPA/DSO Extreme Optics and Imaging (EXTREME) Program award no. HR00111720032

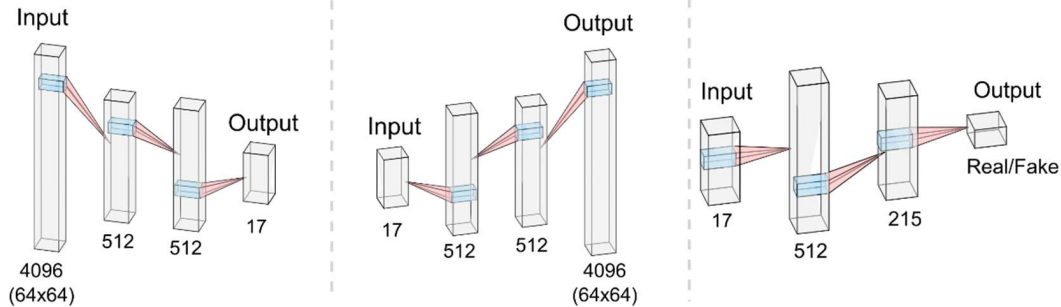
**Figure A1. Structure of the c-AAE network**

### Appendix 1: c-AAE for design production

AAE consists of three coupled neural networks: the encoder, decoder/generator, and discriminator. Figure A1 shows a detailed description of the neural networks.

*Encoder:* The encoder takes a 4096-dimensional vector (that corresponds to a 64 × 64 binary design pattern) as an input. We use two fully-connected layers as the hidden layers of the encoder and a 17 neuron as an output layer of the encoder so that each of the hidden layers has 512 neurons. The last two elements are the geometrical parameters of the unit cell used as a conditional label, while the rest is a coordinate of the 15D compressed design space. For hidden layers, the rectified linear unit (ReLU) activation function is used, and one batch normalization layer is coupled to the second linear layer.

*Decoder:* The decoder has the same architecture as the encoder but with the reversed sequence. The decoder generates a 4096-element output vector based on 17-dimensional input (15D coordinate vector + 2 conditional labels). For the output layer, we use the $\tanh$ activation function.

*Discriminator*: The discriminator takes a 17-dimensional latent vector as an input and binary perform classification (fake/real), so the output is one neuron. Here we have used 2 hidden liner layers with 512 and 256 neurons. The activation function for two hidden layers is the ReLU and for the output layer is the sigmoid function.

### Appendix 1: c-VGGnet structure

CNN takes 64 × 64 image of the design as an input and passes it through three hidden layers, which consist of convolutional layers with ReLU activation functions. Each hidden layer is followed by the max. pooling layer, which ensures the down-sampling of the feature maps. The stack of convolutional layers is followed by one fully-connected layer, which is paired with 2 conditional labels corresponding to the unit cell geometrical parameters. The base VGGnet architecture is followed by "linear" activation function with

"mean squared error" loss function for efficiency prediction (regression). A detailed description of the VGGnet is shown in Fig. A2.

| Layer (type) | Output Shape | Param # |
|---|---|---|
| Input | [64,64,3] | |
| conv2d_1 | [64, 64, 32] | 896 |
| activation_1 | [64, 64, 32] | 0 |
| batch_normalization_1 | [64, 64, 32] | 128 |
| max_pooling2d_1 | [21, 21, 32] | 0 |
| dropout_1 | [21, 21, 32] | 0 |
| conv2d_2 | [21, 21, 64] | 18496 |
| activation_2 | [21, 21, 64] | 0 |
| batch_normalization_2 | [21, 21, 64] | 256 |
| conv2d_3 | [21, 21, 64] | 36928 |
| activation_3 | [21, 21, 64] | 0 |
| batch_normalization_3 | [21, 21, 64] | 256 |
| max_pooling2d_2 | [10, 10, 64] | 0 |
| dropout_2 | [10, 10, 64] | 0 |
| conv2d_4 | [10, 10, 128] | 73856 |
| activation_4 | [10, 10, 128] | 0 |
| batch_normalization_4 | [10, 10, 128] | 512 |
| conv2d_5 | [10, 10, 128] | 147584 |
| activation_5 | [10, 10, 128] | 0 |
| batch_normalization_5 | [10, 10, 128] | 512 |
| max_pooling2d_3 | [5, 5, 128] | 0 |
| dropout_3 | [5, 5, 128] | 0 |
| flatten_1 | [3200] +[2] | 0 |
| dense_1 | [1024] | 3277824 |
| activation_6 | [1024] | 0 |
| batch_normalization_6 | [1024] | 4096 |
| dropout_4 | [1024] | 0 |
| dense_2 | [4] | 4100 |
| dense_3 | [1] | 5 |
| Total params: | | 3,565,449 |

**Figure A2. C-VGGnet structure.**